\documentclass[conference]{IEEEtran}
\IEEEoverridecommandlockouts
\usepackage{cite}
\usepackage{amsmath,amssymb,amsfonts}
\usepackage{algorithmic}
\usepackage{graphicx}
\usepackage{textcomp}
\usepackage{xcolor}

\usepackage{authblk}

\def\BibTeX{{\rm B\kern-.05em{\sc i\kern-.025em b}\kern-.08em
    T\kern-.1667em\lower.7ex\hbox{E}\kern-.125emX}}
\begin{document}

\title{QuCNN : A Quantum Convolutional Neural Network with Entanglement Based Backpropagation}
 
\author[1]{Samuel Stein\thanks{Samuel.stein@pnnl.gov}}
\author[2]{Ying Mao}
\author[1]{James Ang}
\author[1]{Ang Li\thanks{Ang.li@pnnl.gov}}
\affil[1]{Pacific Northwest National Laboratory}
\affil[2]{Fordham University}

\maketitle

\begin{abstract}
Quantum Machine Learning continues to be a highly active area of interest within Quantum Computing. Many of these approaches have adapted classical approaches to the quantum settings, such as QuantumFlow, etc. We push forward this trend, and demonstrate an adaption of the Classical Convolutional Neural Networks to quantum systems - namely QuCNN. QuCNN is a parameterised multi-quantum-state based neural network layer computing similarities between each quantum filter state and each quantum data state. With QuCNN, back propagation can be achieved through a single-ancilla qubit quantum routine.  QuCNN is validated by applying a convolutional layer with a data state and a filter state over a small subset of MNIST images, comparing the backpropagated gradients, and training a filter state against an ideal target state. 

\end{abstract}


\section{Introduction}

Quantum computing is poised to provide computational speedups that classical computing could never feasibly attain. With continued quantum system development and applications in domains such as Quantum Chemistry, Quantum Simulation, and Quantum Machine Learning \cite{biamonte2017quantum,national2019quantum,bertels2021quantum,preskill2018quantum}, the potential of Quantum Computing continues to grow. With high level algorithm development and low level system design, we continue to improve the current state of the art. Superconducting quantum processors of size 127 qubits have been released by IBM \cite{chow2021ibm}, and 22 qubit trapped ion quantum processors by IonQ \cite{widdows2022near}, all with continued improvements in overall qubit quality.
Quantum Machine Learning has been a field of continued interest within quantum computing, with hopes of applications to adapting classical machine learning success to the quantum setting. However, to really demonstrate quantum advantage for QML, the data that is being processed should be quantum data. For example, through the use of QRAM or Quantum Simulation \cite{di2020fault,brown2010using}. Quantum Machine Learning \cite{cong2019quantum,jiang2021co,stein2022quclassi,stein2021qugan} has borrowed motivation from classical machine learning over the last years, attempting to mimic the classically successful techniques through a quantum routine (QuClassi, QuantumFlow and Convolutional Quantum network \cite{cong2019quantum,jiang2021co,stein2022quclassi}). 
Classical machine learning requires little motivation, and has seen wide spread and ubiquitious success \cite{nayak2017impacts,crowston2019impacts}. Within deep learning, the extremely successful design of a Convolutional Neural Network \cite{o2015introduction} saw an explosion in performance within multiple domains \cite{tabian2019convolutional,hershey2017cnn,krizhevsky2017imagenet}. Convolutional neural networks perform high level feature extraction using the inner product between filters and subsections of data which convolutionally evolve traversing the data. 
In this paper, we propose a natural adaption of classical convolutional neural network techniques to quantum computing through the use of the SWAP test, and demonstrate entanglement style back propagation. We numerically demonstrate the relation, and similarity between the implementation of QuCNN and classically implemented convolutional layers, and finally demonstrate the learning ability of parameterised unitary layers in learning pre-trained solution filters. In this work, a QuCNN layer is applied to a small training sample on MNIST, illustrating the ability to perform a convolutional operation in the quantum setting.

\section{Background}
\subsection{Quantum Computing}

Quantum computing adopts classical computing techniques such as bit representation and combines it with quantum mechanical phenomena such as entanglement and superposition \cite{rieffel2000introduction,kaye2006introduction}. In classical computing, data is represented as either 1 or 0, whereas in quantum computing we represent data as $|1\rangle$,$|0\rangle$, or $\alpha|0\rangle + \beta|1\rangle$, a superposition of both. The coefficients $\alpha$ and $\beta$ represent the probability that the data is in the respective values state. Expanding on this, multiple bits are represented by the state $|\psi\rangle = \alpha|00...0\rangle + \beta|00..1\rangle + ... + \omega|11...1\rangle$, where $2^n$ coefficients represent the quantum state. Finally, quantum computing exposes the computational potential of quantum entanglement. Quantum entanglement is most easily understood via the CNOT gate, which transforms the quantum state of $|00\rangle+|01\rangle$ into $|00\rangle+|11\rangle$. In this transformation, the first bit is flipped only if the second bit is in the $|1\rangle$ state. This exposes computational potential to quantum computers and has no classical analogue.

\subsection{Convolutional Neural Networks}

Convolutional Neural Networks perform high level feature extraction from data that exhibits spatial relations \cite{jogin2018feature}. Images are a prime example of where spatially related data exists, where pixels provide context to pixels around them, and hence information. The convolutional layer is characterised by having a set of filter banks, each of some tensor shape. Each filter independently moves and performs the inner product over the data in a pre-described way. This returns a single value representing the similarity between the filter and the data at that location. These layers are optimised similar to all other neural network layers, through backpropagation over some optimization function.

\section{Related Work}

Adapting classical machine learning techniques to quantum systems is an active area of research, with architectures such as Quantum Convolutional Neural Networks, QuantumFlow, QuGAN and QuClassi. 

Quantum Convolutional Neural Networks \cite{cong2019quantum} - QCNN - adapts the classical notion of spatial data encoding, and adapts it to quantum machine learning techniques. QCNN makes use of dual qubit unitaries and mid-circuit measurement to perform information down pooling, to which decisions can be inferred. This is compared with an opposite direction traversal of a MERA network.

QuClassi \cite{stein2022quclassi} proposes a state based detection scheme, borrowing from classical machine learning approaches of training "weights" to represent classifier states. These states each represent a probability of belonging to the states respective class, which generates output layers synonymous with classical classification network outputs. 

QuantumFlow \cite{jiang2021co} attempts to mimic the transformations undergone in classical neural networks, and attempts to accomplish a similar transformation as the classical $y = f(x^Tw + b)$. This is accomplished via the usage of phase flips, accumulation via a hadamard gate accompanied with an entanglement operation. QuantumFlow demonstrates the advantage of batch normalisation, showing notable performance improvements when normalising quantum data to reside around the XY plane, rather than clustering around either the $|1\rangle$ or $|0\rangle$ point. Furthermore, QuantumFlow demonstrates the reduced parameter potential of quantum machine learning, illustrating a quantum advantage.

Chen 2022 \cite{chen2022quantum} makes use of a quantum convolutional network to perform high energy physics data analysis. In the paper, they present a framework for encoding localised classical data, followed by a fully entangled parameterised layer to perform spatial data analysis. Their numerical analysis demonstrates the promise of quantum convolutional networks.

Notably, all of these works have taken a classical machine learning technique, and adapted it in some form to the quantum setting. We aim to accomplish the same with QuCNN, adapting the convolutional filter operation.

\section{QuCNN}

In this section, we walk through the adaption of a classical convolutional filter operation to QuCNN. We further go on to demonstrate a quantum-implemented backpropagation algorithm allowing for an almost entirely quantum routine to compute the $\frac{dL}{d\theta_i}$ gradient, where $\theta_i$ is the layer weight. 

\subsection{QuCNN Layer Architecture}

Classical convolutional neural network's learn a set of feature maps for local pattern recognition over a trained data set. This operation is characterised by the Convolution (Conv) operation. One Convolution Operation comprises of a filter $F$, and an input $X$, and performs Conv(X,F). This is described by the convolution operation outlined in \ref{eqn:conv}

\begin{equation}
    y_{ij} = \sum_{k=1}^{HH}\sum_{l=1}^{WW}w_{kl}x_{si+k-1,sj+l-1}
    \label{eqn:conv}    
\end{equation}

Importantly, $wx^{'}$ is equivalent to the dot product between two vectors $w$ and $x$. A comparable computation is realised in quantum computing through the SWAP test algorithm, which computes the equation outlined in Equation \ref{eqn:swap}, with error $O(\frac{1}{\epsilon^2})$. Given sufficient samples, the SWAP test is an unbiased estimator of the inner product squared.
\begin{equation}
    SWAP(Q_0,|\psi\rangle,|\phi\rangle) = P(M|Q_0\rangle=0) = \frac{1}{2} + \frac{1}{2}|\langle\psi|\phi\rangle|^2
    \label{eqn:swap}
\end{equation}

Given this operation, we can perform similar convolutional operations by performing the outlined Formula in \ref{eqn:conv_filter}, with i filters:
\begin{equation}
    y_{ij} = SWAP(|\Psi_{i}\rangle,|X\rangle_{si:si+k,sj:sj+l})
    \label{eqn:conv_filter}    
\end{equation}
where the statevector describing $|X\rangle_{si:si+k,sj:sj+l}$ has the same dimensionality, and hence number of qubits, of $|\Psi_{i}\rangle$. The forward operation produces a similar output to a classical convolutional operation, whereby we are computing the squared real inner product of two state vectors instead of the inner product of two vectors.

\begin{figure}
    \centering
    \includegraphics[width=0.5\textwidth]{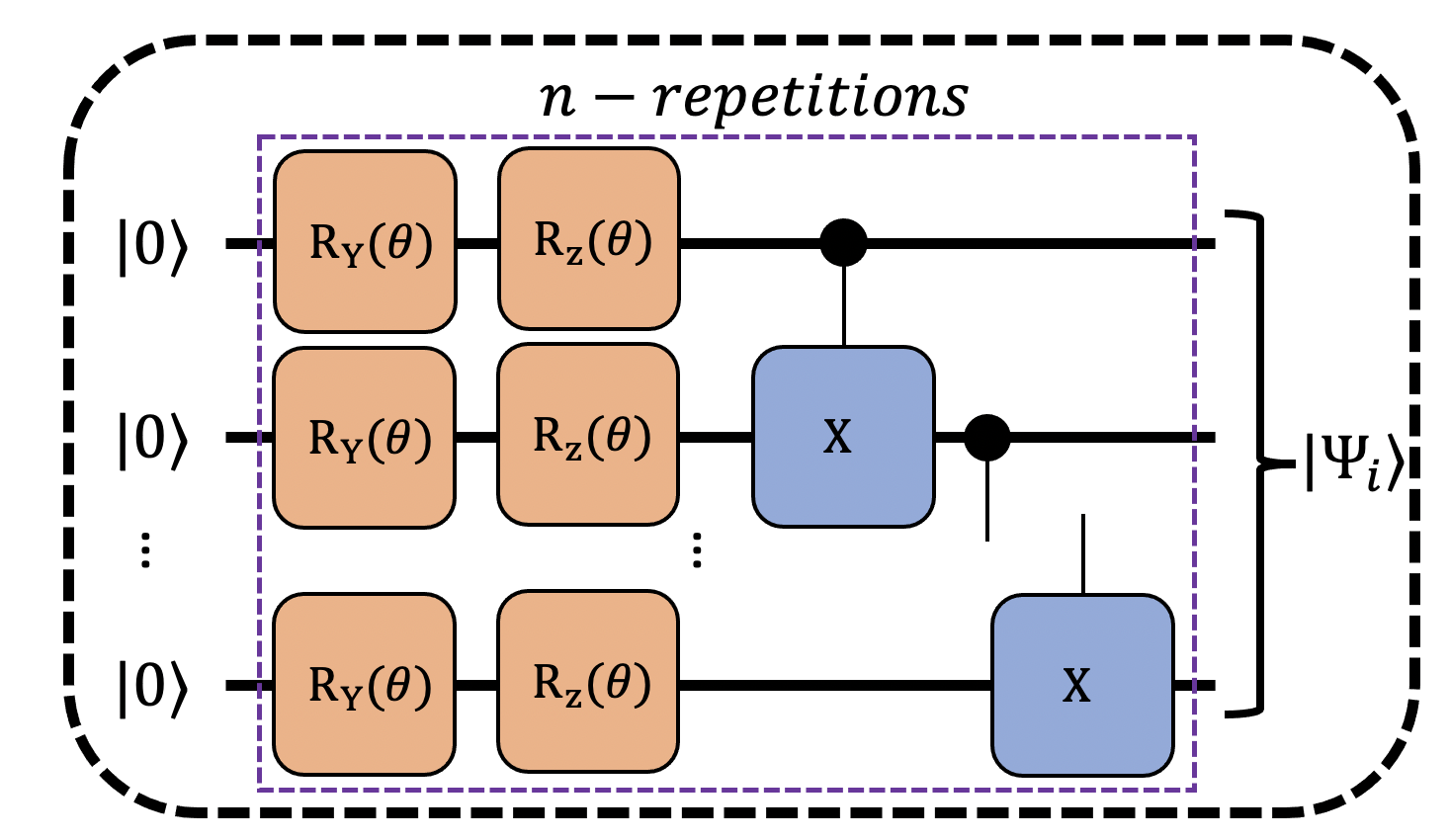}
    \caption{Single convolutional filter layer structure. n-repetitions represents the number of parameterised layers comprising the trainable filter. Each layer is comprised of a complete Y-axis rotation, complete Z-axis rotation, followed by a linear entanglement scheme.}
    \label{fig:layer_struc}
\end{figure}

In classical convolutional networks, the convolutional filter is a tensor of varying activation, all of which are independently optimised according to some loss function. With a quantum state prepared via any ansatz, there is no way to control one state amplitude's magnitude without changing another amplitude's magnitude. This is due to the square norm requirement of quantum states. Therefore, we optimise each quantum state filter similar to the optimization procedure of a variational quantum algorithm such as a variational quantum eigensolver \cite{cerezo2021variational,du2022efficient,farhi2014quantum,kandala2017hardware}. This is visualised in Figure \ref{fig:layer_struc}, where n layers represents the number of parameterised layers describing the quantum state. Each filter maintains its own independent set of $\theta$'s, where each filter attempts to learn its own feature set.

The QuCNN architecture is applicable to purely quantum data, that might be accessed via QRAM or other sources. However, QuCNN can operate on classical data, once the classical data is translated into a quantum state prior to model induction or training. Although computationally expensive, this is a pre-processing step. Within this paper, we utilize a classical-to-quantum encoding technique. Utilizing $log_2(n)$ encoding \cite{jiang2021co}, an input data point is broken up into spatially related data clusters, with a pattern defined by parameters such as stride, filter size etc., and translated into a group of equivalent amplitude encoded state vectors $[|X\rangle_1,|X\rangle_2,|X\rangle_3,...,|X\rangle_n$]. Each filter in the convolutional layer is represented by a circuit ansatz comprised of Pauli-\{X,Z\} rotations and CNOT-entanglement parameterized by $[\theta]$ values - the weight equivalents -  across $log_2(n)$ qubits.

\begin{figure*}
    \centering
    \includegraphics[width=0.8\textwidth]{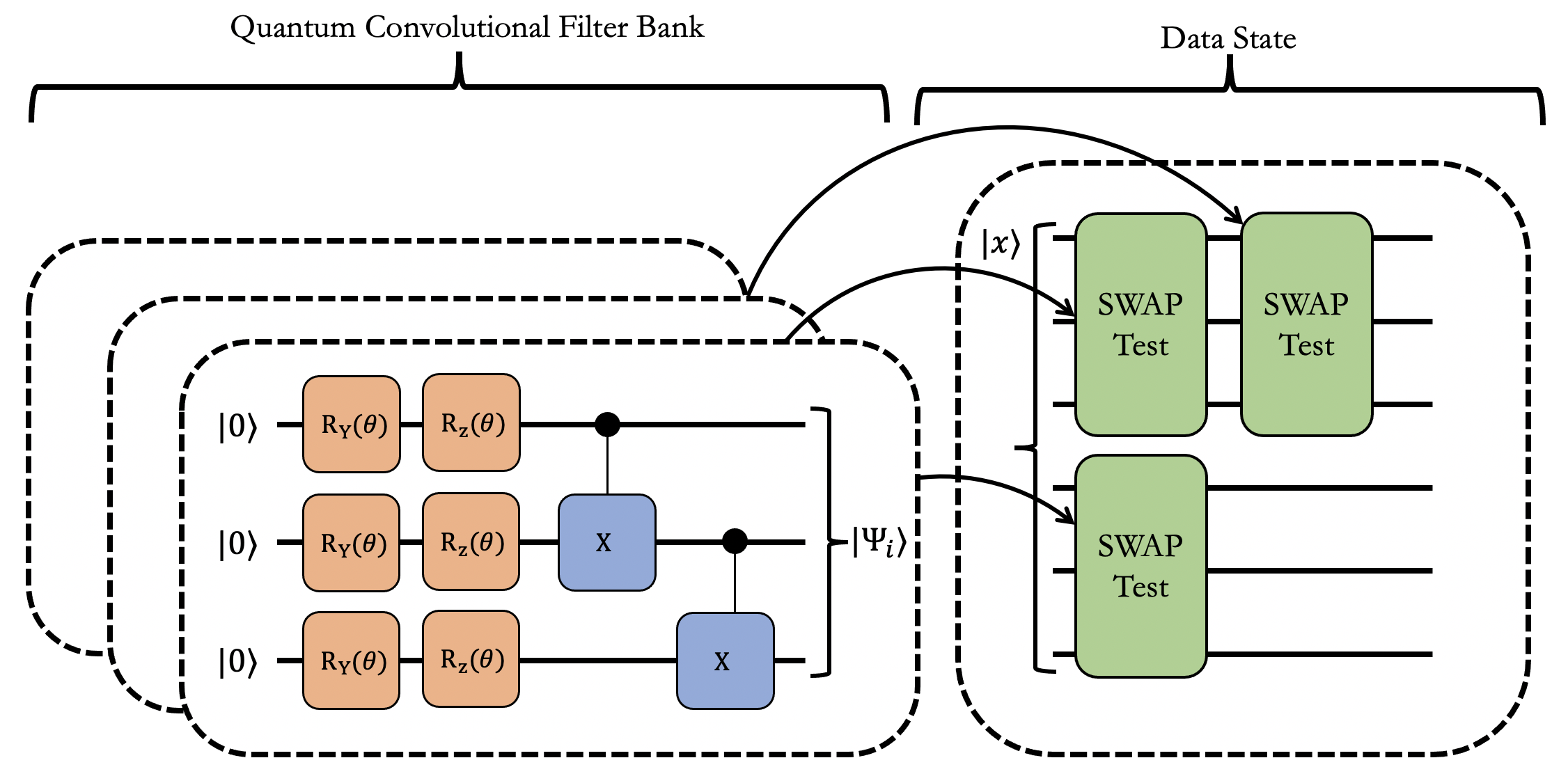}
    \caption{High level application of a Quantum Convolutional Neural Network. Each filter is prepared independent of each other, and a SWAP operation convolves over the quantum state. Notably, SWAP tests are performed in series, with each state operating sequentially. Similar to classical convolutional operations, the SWAP Test moves over the quantum state according to a pre-described movement, similar to Stride.}
    \label{fig:conv_op}
\end{figure*}

The convolutional operation is described in Figure \ref{fig:conv_op}, where each trained filter applies a SWAP test operation to a set of qubits on the data state. The data state is practitioner described, and can be a QRAM access, or a prepared state.

\subsection{Quantum Backpropagation}

In classical layers, the task of optimizing a layer's values of $F_{i,j}$ is characterized by the differential $\frac{dL}{dF}$. Utilizing the chain rule we obtain $\frac{dL}{dF} = \frac{dL}{dO}\frac{dO}{dF}$, whereby $\frac{dL}{dO}$ is the back propagated values from the layer in front of the convolutional layer.  Optimizing classical convolutional layers has a seamless reduction whereby one performs the Conv operation with the back propagated vector over the input, which gives the gradients. For a quantum system there is no manipulating single amplitude magnitudes in state vectors without affecting other amplitudes. Therefore, we can not do similarly styled back propagation. We still have access to $\frac{dL}{dO}$, and want to compute $\frac{dO}{d\theta}$. As $\frac{dL}{d\theta} = \frac{dO}{d\theta}\frac{dL}{dO}$, we make use of the parameterized difference formula over parameterized pauli gates to attain $\frac{dO}{d\theta}$. This is characterized by Equation \ref{eqn:diff}, and the $\frac{pi}{2}$ can be tuned according to the problem setting.
\begin{equation}
    \frac{df}{d\theta} = \frac{1}{2}({f(\theta+\frac{\pi}{2}}) - {f(\theta-\frac{\pi}{2}}))
    \label{eqn:diff}
\end{equation}

Since the swap test measures only one qubit, we are computing simply the output expectation of one ancilla qubit, $f(\theta)=\mathbf{E}(\text{M}_Z|Q_0\rangle) = P(M_Z|Q_0\rangle=0) = \frac{1}{2} + \frac{1}{2}|\langle X_j|\Psi(\theta)\rangle|^2$ where $M$ is a measurement in the Pauli-Z basis after the SWAP test has been performed, with the control qubit $Q_0$. Expanding on Equation \ref{eqn:diff}.

\begin{multline}
    \frac{dO_j}{d\theta_i} =\frac{1}{2} ( SWAP(|\Psi[\theta_{i}+\frac{\pi}{2}]\rangle,|X_j\rangle) -  \\  SWAP(|\Psi[\theta_{i}-\frac{\pi}{2}]\rangle,|X_j\rangle))
\end{multline}

\begin{multline}
     \frac{dO_j}{d\theta_{i}} =\frac{1}{2}(\frac{1}{2} + \frac{1}{2}|\langle X_j|\Psi(\theta_i + \frac{\pi}{2})\rangle|^2 -  \frac{1}{2} - \frac{1}{2}|\langle X_j|\Psi(\theta_i - \frac{\pi}{2})\rangle|^2) \\
     \frac{dO_j}{d\theta_{i}} =\frac{1}{4}(|\langle X_j|\Psi(\theta_i + \frac{\pi}{2})\rangle|^2 - |\langle X_j|\Psi(\theta_i - \frac{\pi}{2})\rangle|^2) \\
     \frac{dO}{d\theta_i} =  \sum_j^{W\text{x}H} \frac{1}{4}(|\langle X_j|\Psi(\theta_i + \frac{\pi}{2})\rangle|^2 - |\langle X_j|\Psi(\theta_i - \frac{\pi}{2})\rangle|^2)\\
     \frac{dL}{d\theta_i} = \sum_j^{W\text{x}H} \frac{1}{4}\frac{dL}{dO_j}(|\langle X_j|\Psi(\theta_i + \frac{\pi}{2})\rangle|^2 - |\langle X_j|\Psi(\theta_i - \frac{\pi}{2})\rangle|^2) \\ 
     \label{eqn:proof_desc}
\end{multline}

For Equation \ref{eqn:proof_desc} to operate on a quantum processor, scaling by $\frac{dL}{dO_j}$ must be applied on the quantum processor. We provide that routine through the use of a controlled entangled ancilla qubit. 

Once the SWAP test has been performed between two states $|\phi \rangle$ and $|\psi\rangle$, the ancilla qubit remains in the state in Equation \ref{eqn:swap_state}.

\begin{multline}
    |Q_{SWAP}\rangle = \frac{1}{2}|0\rangle(|\phi,  \psi \rangle + | \psi , \phi \rangle)  + \frac{1}{2}|1\rangle(|\phi,  \psi \rangle - |\psi , \phi \rangle )
    \label{eqn:swap_state}
\end{multline}
We introduce an ancilla qubit $Q_\text{Anc}$, and CNOT $Q_{SWAP}$ with $Q_\text{Anc}$ as the control qubit.
\begin{equation}
    |Q_\text{Anc}\rangle = \alpha|0\rangle + \beta|1\rangle
    \label{eqn:q_anc_sv}
\end{equation}
\begin{multline}
    |Q_{Anc},Q_{SWAP}\rangle = \frac{\alpha}{2}(\langle \phi,  \psi | + \langle \psi , \phi| )|00\rangle + \\
    \frac{\alpha}{2}(\langle \phi,  \psi | - \langle \psi , \phi |)|01\rangle + \\
    \frac{\beta}{2}(\langle \phi,  \psi | + \langle \psi , \phi |)|10\rangle + \\
    \frac{\beta}{2}(\langle \phi,  \psi | - \langle \psi , \phi |)|11\rangle\\
\end{multline}
$Q_{SWAP}$ is entangled with $Q_{Anc}$ through a CNOT gate, with $Q_{Anc}$ as the control qubit.
\begin{multline}
    CNOT(|Q_{Anc}\rangle,|Q_{SWAP}\rangle) = \frac{\alpha}{2}(\langle \phi,  \psi | + \langle \psi , \phi )|00\rangle + \\
    \frac{\alpha}{2}(\langle \phi,  \psi | - \langle \psi , \phi| ) |01\rangle + \\
    \frac{\beta}{2}(\langle \phi,  \psi | - \langle \psi , \phi| ) |10\rangle + \\
    \frac{\beta}{2}(\langle \phi,  \psi | + \langle \psi , \phi| ) |11\rangle\\
\end{multline}
In its current state, our new $P(\text{M}_Z|Q_{Anc}\rangle)$ changes.
\begin{multline}
    P(\text{M}_Z|Q_{SWAP}\rangle = \\ (\frac{\alpha}{2}(\langle \phi,  \psi | + \langle \psi , \phi| ) + \frac{\beta}{2}(\langle \phi,  \psi | - \langle \psi , \phi | ))^2
\end{multline}
For ease of representation, we let $\langle \phi, \psi|=X$, and $\langle \psi, \phi|=Y$
\begin{multline}
    P(|Q_{SWAP}\rangle = 0)= (\frac{\alpha}{2}(X + Y) + \frac{\beta}{2}(X - Y ))^2
    \label{eqn:x_y_p0}
\end{multline}
Expanding Equation \ref{eqn:x_y_p0};
\begin{multline}
    P(\text{M}_Z|Q_{Anc}\rangle = \frac{1}{4} \times (\alpha X + \alpha Y +\beta X - \beta Y )^2 \\
    P(\text{M}_Z|Q_{Anc}\rangle = \frac{1}{4} \times (\alpha ^2 X ^2 + \alpha ^2 XY + \alpha \beta X^2 - \alpha \beta XY \\+ \alpha ^2 Y^2 + \alpha \beta XY - \alpha \beta Y^2 + \beta^2X^2 - \beta^2XY + \beta^2Y^2 )
\end{multline}
The inner product of two vectors with themselves is equal to one (i.e. identical), hence we know that $\langle \psi,\phi|\phi,\psi\rangle = 1$ and $\langle \phi,\psi|\psi,\phi\rangle = 1$, and that $X^2=Y^2=1$
\begin{multline}
    P(\text{M}_Z|Q_{Anc}\rangle = \frac{1}{4} \times (\alpha ^2 + \alpha ^2 XY + \alpha \beta  - \alpha \beta XY \\+ \alpha ^2 + \alpha \beta XY - \alpha \beta+ \beta^2- \beta^2XY + \beta^2) \\
    P(\text{M}_Z|Q_{Anc}\rangle = \frac{1}{4} \times (2(\alpha ^2 + \beta^2)  -(\alpha^2 - \beta^2)XY \\
    P(\text{M}_Z|Q_{Anc}\rangle = \frac{1}{2} \times ((\alpha ^2 + \beta^2)  -\frac{(\alpha^2 - \beta^2)}{2}XY )
\end{multline}
State vectors must have a square norm of 1, therefore $\alpha^2 + \beta^2 = 1$ and $\alpha^2=1-\beta^2$.
\begin{multline}
    P(\text{M}_Z|Q_{Anc}\rangle = \frac{1}{2} \times (1  - \frac{(1-2\beta^2)}{2}XY \\
    P(\text{M}_Z|Q_{Anc}\rangle = \frac{1}{2}  - (\frac{1}{2} - \beta^2)|\langle \phi,\psi \rangle|^2   
    \label{eqn:proof_control}
\end{multline}
In Equation \ref{eqn:proof_control}, we have shown that we can control the magnitude of our inner product by a control ancilla qubit. Putting this into our parameterized difference formula, we can now accomplish the chain rule using an ancilla qubit.
\begin{multline}
    \frac{dL}{d\theta_i} = \sum_j^{W\text{x}H} \frac{1}{4}\frac{dL}{dO_j}((\frac{1}{2}-\beta^2)|\langle X_j|\Psi(\theta_i + \frac{\pi}{2})\rangle|^2 \\ - (\frac{1}{2}-\beta^2)|\langle X_j|\Psi(\theta_i - \frac{\pi}{2})\rangle|^2
\end{multline}
We can absorb the $\frac{dL}{dO_j}$ through the $\frac{1}{2}-\beta^2$, therefore giving us the following derivation.
\begin{multline}
    \frac{dL}{d\theta_j} = \sum_j^{W\text{x}H} \frac{dO}{d\theta_j}\frac{dL}{dO} \\ 
    \frac{dL}{d\theta_j} = \sum_j^{W\text{x}H} \frac{dL}{4dO}(f(\theta_j+\frac{\pi}{2}) - f(\theta_j)-\frac{\pi}{2}) \\ 
    \frac{dL}{d\theta_i} = \sum_j^{W\text{x}H} \frac{1}{4}(\overbrace{\frac{dL}{dO_j}|\langle X_j|\Psi(\theta_i + \frac{\pi}{2})\rangle|^2}^{\mathbf{E}(Q_{Anc(\theta_i + \frac{\pi}{2})}} \\ - \overbrace{\frac{dL}{dO_j}|\langle X_j|\Psi(\theta_i - \frac{\pi}{2})\rangle|^2}^{Q_{Anc(\theta_i - \frac{\pi}{2})})}
    \label{eqn:backprop}
\end{multline}

\begin{figure*}
    \centering
    \includegraphics[width=0.85\textwidth]{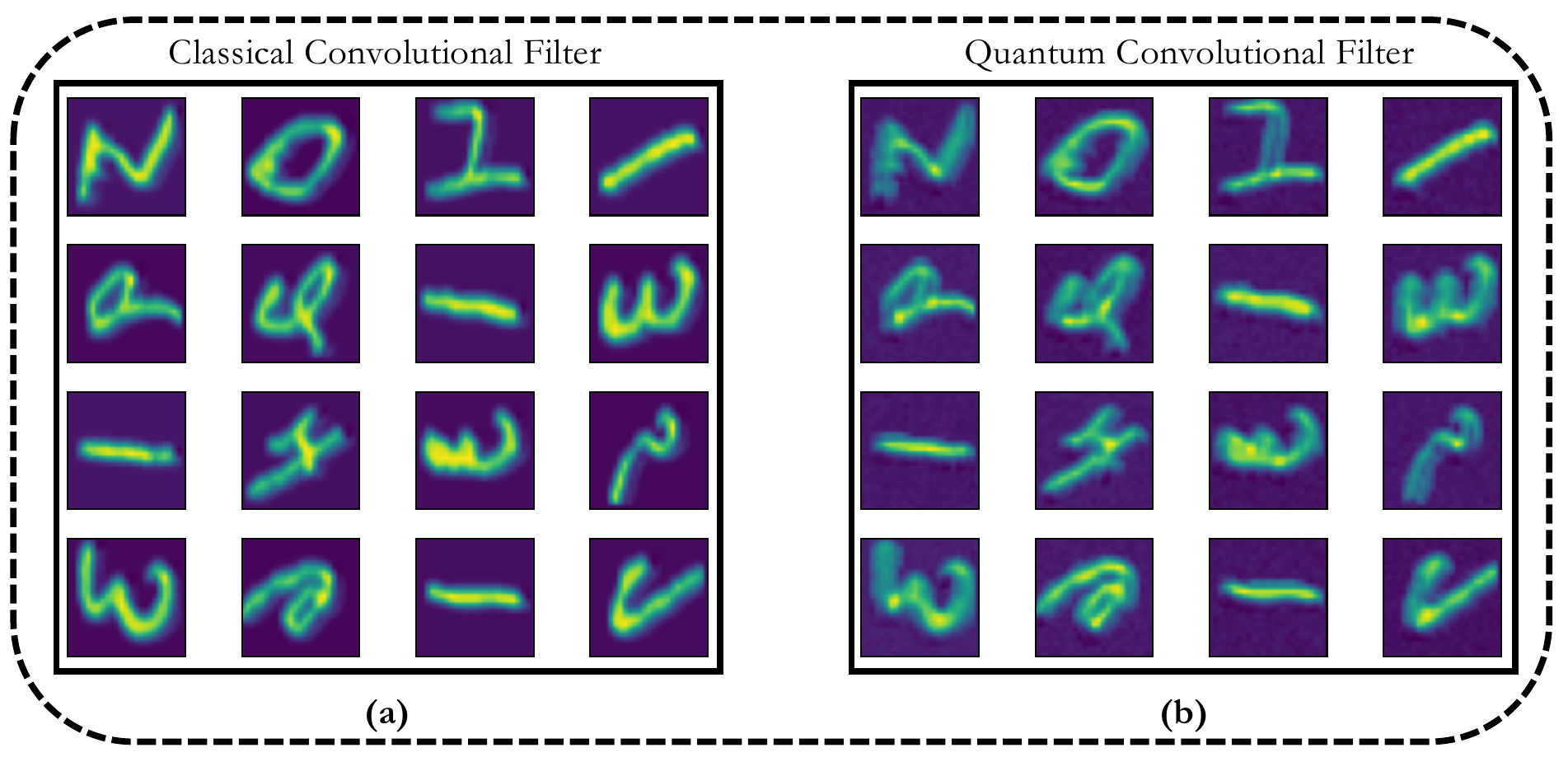}
    \caption{Comparison of a trained classical filter passed through a classical convolutional operation with the filter transformed into the detector state on a quantum state over 16 MNIST images. \textbf{(a)} represents a trained classical convolutional layer passed over 16 MNIST images. \textbf{(b)} represents taking the same trained filter, translating it to a quantum state, and using it as the filter state in the QuCNN algorithm. The MNIST images are transformed into unitary matrices and passed as a unitary gate, with the output being sampled from the SWAP test ancilla qubit.}
    \label{fig:large_results}
\end{figure*}

$b^2$ is bound in the range of $[0,1]$, therefore the range of $\frac{dL}{dO_j}$ is $[-0.5,0.5]$. We can perform a $R_Y$ gate parameterized by $\theta_\beta$ on $Q_{Anc}$ prior to entangling from the state $|0\rangle \xrightarrow[]{} cos(\frac{\theta}{2}|0\rangle - iSin(\frac{\theta_\beta}{2})|1\rangle$. Therefore, $\beta^2 = (-iSin(\frac{\theta_\beta}{2})^2 = sin^2\frac{\theta_\beta}{2}$. Therefore, $\frac{dL}{dO_j} = \frac{1}{2} - sin^2{\frac{\theta_\beta}{2}} \xrightarrow[]{} \theta_\beta = 2\sqrt{\frac{1}{2} - \frac{dL}{dO_j}}$

\section{QuCNN Analysis}

In this section, we put demonstrate the applicability of the QuCNN structure, and how the layer functions when implemented in comparison with classical convolutional layers. Within this, our evaluation comprises taking a classically learnt filter, translating it into an equivalent applicable quantum state, and using it as the convolving filter over 16 MNIST images. The resultant output feature map is compared with the classically generated feature map. We go on to demonstrate that Equation \ref{eqn:backprop}, when applied to a circuit, generates the correct result of $\frac{dL}{d\theta_i}$, as well as train a parameterised ansatz to mimic a learned filters statevector.

Notably, due to complexity of training, we do not train a network with QuCNN. We observe challenges within this, as we believe to be due to barren plataeus. Our evaluation comprises a demonstration of the transfer of a classical technique to quantum, validating both the forward and backward passes, and training a parameterised circuit with a target ideal filter. Future improvements in quantum machine learning training however, are most likely required to train with this layer from end to end.

\subsection{Forward Pass Validation}

\begin{figure}
    \centering
    \includegraphics[width=0.5\textwidth]{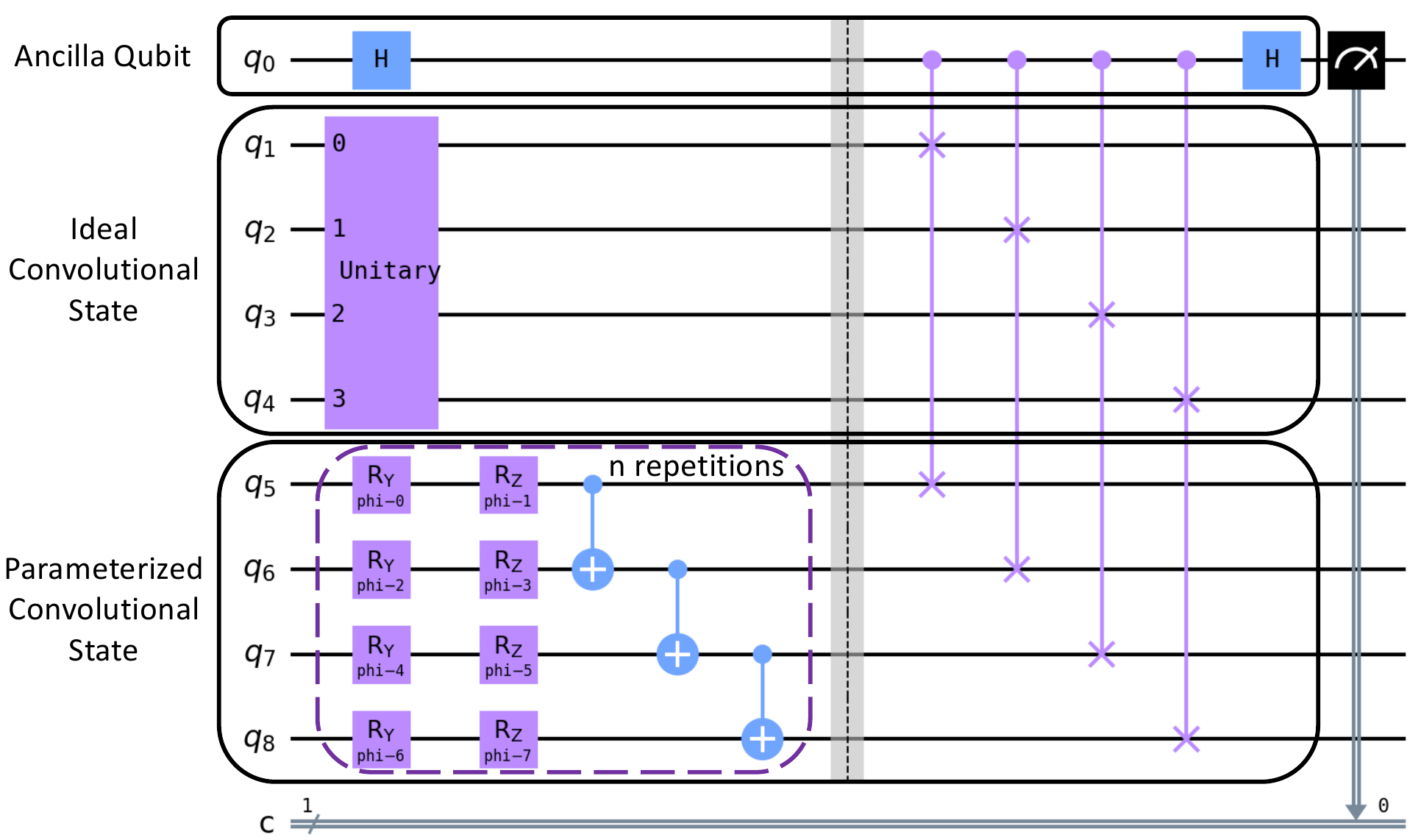}
    \caption{Trainable Convolutional Filter Circuit with training described under Figure \ref{fig:training_state}. Ancilla qubit represents the SWAP Test ancilla qubit, Ideal Conovlutional State the theoretical state learnt classically transformed into a unitary operation, and Parameterized Convolutional State an untrained state attempting to learn to mimic the ideal convolutional state.}
    \label{fig:state_circuit}
\end{figure}

In this section, we go on to demonstrate that QuCNN is able to mimic the forward pass outputs of a classical convolutional neural network. In this, we train a  single classical convolutional layer comprising of one 4x4 filter within a classical neural network. Once trained, we take the 4x4 filter, set the norm to 1, thereby making the filter able to be represented by a quantum state vector. We generate a unitary matrix that transforms a quantum state from the ground state into this quantum state. This is demonstrated over the Ideal Convolutional State block represented in Figure \ref{fig:state_circuit}. We take 16 MNIST images, and decompose them into 625 4x4 portions, defined by the convolutional operation with a stride of 1. Each portion is processed through Singular Value Decomposition, generating a unitary matrix that represents the transformation required to generate a quantum state representing the small 4x4 portion. This process is done in QuantumFlow \cite{jiang2021co}. We perform the loading process over each convolutional section on the MNIST image, and given a unitary gate operation that transforms the state into a unit state representing each section, we SWAP test the aforementioned filter state with the data state. We measure the ancilla qubit with 10000 samples, and thereby the fidelity between the two states. We plot the resultant fidelities as a 25x25 grid, similar to the 25x25 grid returned from the classical network. This is visualised in Figure \ref{fig:large_results}. 

We observe that the Quantum Convolutional operation is able to mimic the classical operations outputs. As can be seen in Figure \ref{fig:large_results}, the Quantum Convolutional Filter outputs replicate the Classical Convolutional Filter outputs with noisy edges. This can be attributed to the sampling error mentioned above. Although this is not a novel result, this demonstrates the similarity between the quantum convolutional operation of QuCNN, and the classical counterpart. Pattern detection within domains such as Quantum Sensing are extremely important domains, with little information as to what might be the most applicable solution, with hopes that Quantum Machine Learning might offer data discovery opportunities.

\subsection{Backpropagation Validation}

\begin{figure}
    \centering
    \includegraphics[width=0.5\textwidth]{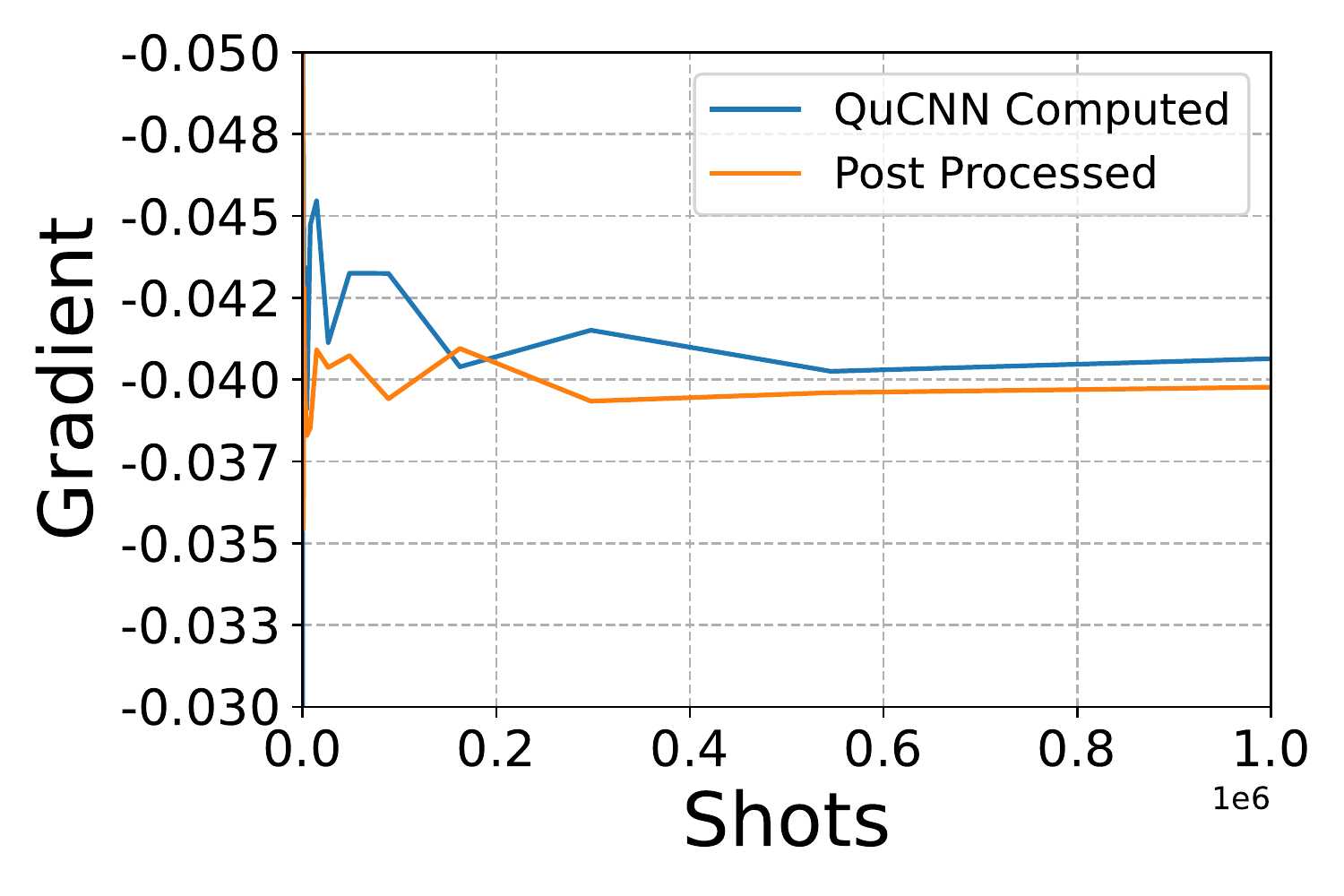}
    \caption{Comparison of QuCNN quantum-integrated back propagation against non-integrated approach. QuCNN Computed represents gradients returned via Equation \ref{eqn:backprop}, whereas post processed uses the regular approach, with no ancilla chained qubit described in Equation \ref{eqn:diff}}
    \label{fig:backprop_val}
\end{figure}

In the following section, we demonstrate the validity of using an ancilla qubit to move the backpropagated value into the quantum routine. This is demonstrated in Figure \ref{fig:backprop_val}. In this figure, we generate a convolutional filter, and SWAP test it with a maximally-mixed state. We perform Equation \ref{eqn:backprop}, with a backpropagated value of $0.3$. We compare the resultant $\frac{dL}{d\theta_i}$ gradient with the parametric difference method (i.e. post processed). Notably, there is naturally some error involved due to the sampling nature of computing quantum gate gradients via sampling.  The results of this are visualized in Figure \ref{fig:backprop_val}, whereby we observe that given sufficient shots, the gradient observed from QuCNN and Post Processed converge to similar values. Lower shot numbers are much noisier, attributed to the sampling error of the SWAP test.

\begin{figure}
    \centering
    \includegraphics[width=0.5\textwidth]{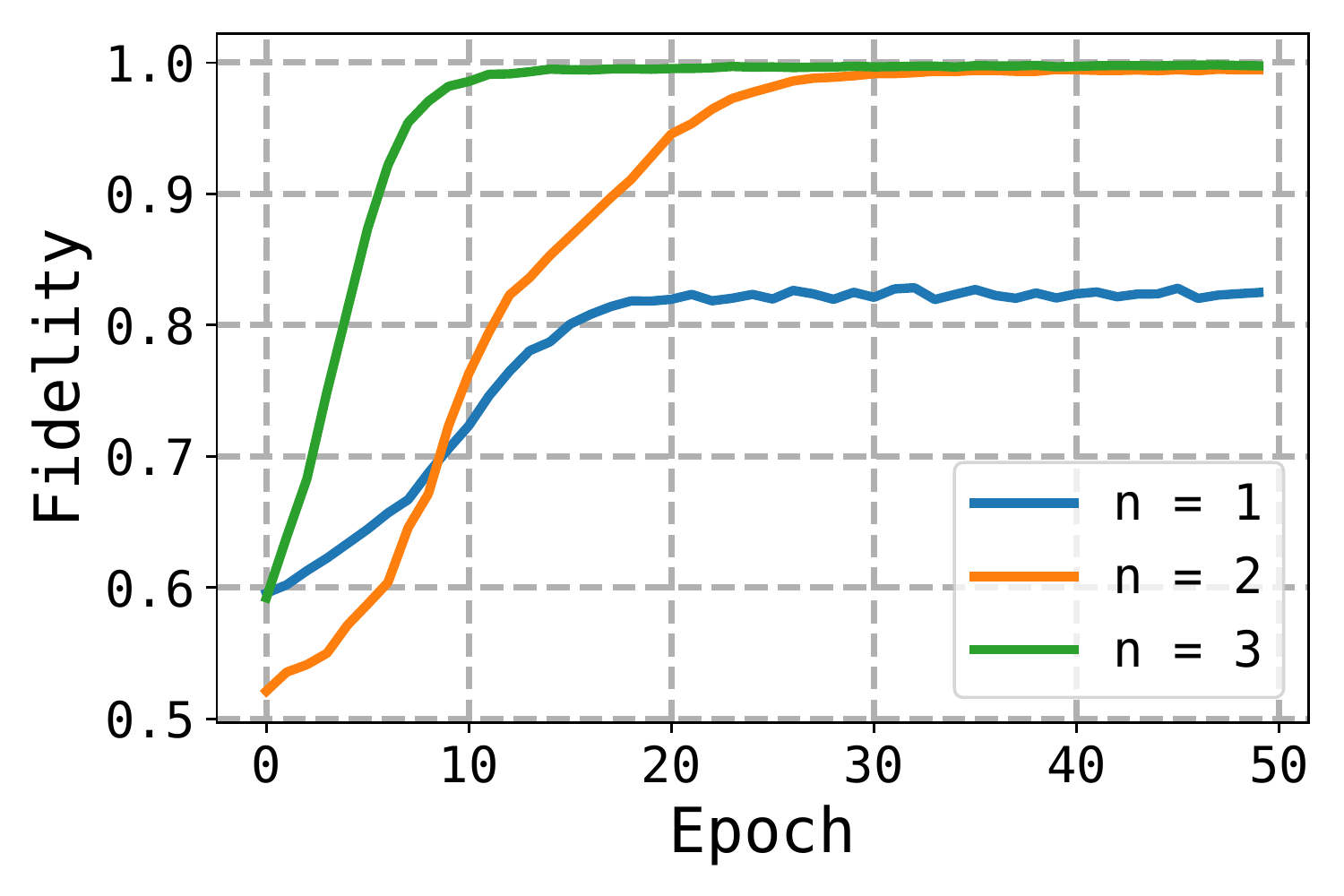}
    \caption{Training of 3 n value states depicted in Figure \ref{fig:state_circuit}. Training is optimising the untrained parameterized state to mimic the ideal convolutional state which is trained pre-induction on a classical network.}
    \label{fig:training_state}
\end{figure}

\subsection{State Learning Ability}

In Figure \ref{fig:large_results}, we demonstrate that the QuCNN operation is able to perform similarly to classical convolutional filters. However, this was shown on a pre-trained state generated via an ideal unitary. Quantum computers can not implement arbitrary matrices, and require gate decomposition to basis gates. In this section, we demonstrate that these states are in fact expressable as a shallow circuit comprised of n=[1-3] with gates that run on real quantum hardware, as expressed in Figure \ref{fig:layer_struc}. We optimize our parameters for each layer to maximise the fidelity between the parameterised convolutional state and the ideal convolutional state. The learning performance is plotted in Figure \ref{fig:training_state}. 

We observe in this figure that the ideal trained unitaries are not overly complex for the modest parameterised circuits demonstrated in Figure \ref{fig:layer_struc}. These circuits are not domain specific, and are simplistic structures for training variational circuits. The performance observed in Figure \ref{fig:training_state} motivates us to believe these filters are indeed learnable, and do not require overly complex ansatz to generate performant filters.

\section{Conclusion}
In this paper we present an adaption of the classical convolutional layer to  quantum convolutional layers. Furthermore, we have demonstrated a technique for moving the backpropagation portion of training layers in neural networks to be predominantly on the quantum processor. We validate the approaches by applying them. Mimicking classical machine learning techniques on quantum processors, whilst not providing immediate computational advantage, sets the stage for future quantum computing developments such as QRAM, improved Quantum Processor performance, and beyond.

\section*{Acknowledgments}
This material is based upon work supported by the U.S. Department of Energy, Office of Science, National Quantum Information Science Research Centers, Co-design Center for Quantum Advantage ($C^2QA$) under contract number DESC0012704. We would like to thank the PNNL operated IBM-Q Hub supported by DOE Office of Science, Advanced Scientific Computing Research program. This research used resources of the National Energy Research Scientific Computing Center (NERSC), a U.S. Department of Energy Office of Science User Facility located at Lawrence Berkeley National Laboratory, operated under Contract No. DE-AC02-05CH11231 using NERSC award ERCAP0022228. The Pacific Northwest National Laboratory is operated by Battelle for the U.S. Department of Energy under Contract DE-AC05-76RL01830. 

\bibliographystyle{plain}
\bibliography{references}

\end{document}